\documentclass[cits]{PoS}

\usepackage{amsmath}
\usepackage{graphicx}

\title{Hadron physics and dynamical chiral symmetry breaking\thanks{
We are grateful to the organisers for the opportunity to be involved in this workshop.
Parts of the original material described in this contribution were drawn from collaborations and discussions with A.~Bashir, S.\,J.~Brodsky, C.~Chen, H.~Chen, I.\,C.~Clo\"et, B.~El-Bennich, X.~Guti\'{e}rrez-Guerrero, R.\,J.~Holt, Y.-x.~Liu, V.~Mokeev, T.~Nguyen, S.-x.~Qin, H.\,L.\,L.~Roberts, R.~Shrock and P.\,C.~Tandy.
This work was supported by
U.\,S.\ Department of Energy, Office of Nuclear Physics, contract no.~DE-AC02-06CH11357.
}}

\ShortTitle{Hadron physics and dynamical chiral symmetry breaking}

\author{Lei Chang\\
        Physics Division, Argonne National Laboratory, Argonne, Illinois 60439, USA\\
        E-mail: \email{lei.chiong@gmail.com}}

\author{\speaker{Craig D.~Roberts}\\
        Physics Division, Argonne National Laboratory, Argonne, Illinois 60439, USA;\\
        Department of Physics, Illinois Institute of Technology, Chicago, Illinois \\
        E-mail: \email{cdroberts@anl.gov}}

\author{David J.~Wilson\\
        Physics Division, Argonne National Laboratory, Argonne, Illinois 60439, USA\\
        E-mail: \email{djwilson@anl.gov}}

\abstract{
Physics is an experimental science; and a constructive feedback between theory and extant and forthcoming experiments is necessary if an understanding of nonperturbative QCD is to be achieved.  The Dyson-Schwinger equations connect confinement with dynamical chiral symmetry breaking, both with the observable properties of hadrons, and hence can plausibly provide a means of elucidating the empirical content of strong QCD.  We illustrate these points via comments on: in-hadron condensates; dressed-quark anomalous chromo- and electro-magnetic moments; the self-limiting magnitudes of such moments and pion-loop contributions to the gap equation; deep inelastic scattering; the spectra of mesons and baryons; the critical role played by hadron-hadron interactions in producing these spectra; and nucleon elastic and transition form factors.
}

\FullConference{International Workshop on QCD Green's Functions, Confinement and Phenomenology\\
		 5-9 September 2011\\
		 Trento, Italy}

\begin{document}

\section{Universal Truths}
To set the scene, we recapitulate some basic facts.
The hadron spectrum, and hadron elastic and transition form factors provide unique information about the long-range interaction between light-quarks and the distribution of a hadron's characterising properties amongst its QCD constituents.
Dynamical chiral symmetry breaking (DCSB) is the most important mass generating mechanism for visible matter in the Universe, and this means the Higgs mechanism is (almost) irrelevant to light-quarks.  Notably, this is acknowledged by the Higgs-hunters at CERN who, in announcing inconclusive evidence of a Higgs sighting, stated: ``\emph{The Higgs field is often said to give mass to everything. That is wrong. The Higgs field only gives mass to some very simple particles. The field accounts for only one or two percent of the mass of more complex things like atoms, molecules and everyday objects, from your mobile phone to your pet llama. The vast majority of mass comes from the energy needed to hold quarks together inside atoms}.''
The running of the quark mass, illustrated in Fig.\,\ref{fig:Fig1}, entails that calculations at even modest $Q^2$ require a Poincar\'e-covariant approach.   Furthermore, covariance plus the momentum-dependence of the interaction in QCD together guarantee the existence of quark orbital angular in a hadron's rest-frame wave function.
Confinement is expressed through a violent change in the analytic structure of the propagators for coloured particles; its presence can be read from a plot of a states' dressed-propagator; and it is intimately connected with DCSB \cite{Krein:1990sf}.
We now illustrate how the complex of Dyson-Schwinger equations (DSEs) has been employed to elucidate aspects of confinement and DCSB, and their impact on hadron observables.  Fuller explanations may be found in recent reviews \cite{Roberts:2007ji,Holt:2010vj,Chang:2011vu,Bashir:2012}.

\section{Condensates are confined within hadrons}
\label{sec:inmeson}
Dynamical chiral symmetry breaking and its connection with the generation of hadron masses was first considered in Ref.\,\cite{Nambu:1961tp}.  The effect was represented as a vacuum phenomenon.  Two essentially inequivalent classes of ground-state were identified in the mean-field treatment of a meson-nucleon field theory: symmetry preserving (Wigner phase); and symmetry breaking (Nambu phase).  Notably, within the symmetry breaking class, each of an uncountable infinity of distinct configurations is related to every other by a chiral rotation.  This is arguably the origin of the concept that strongly-interacting quantum field theories possess a nontrivial vacuum.

\begin{figure}[t]\begin{minipage}[t]{\textwidth}
\begin{minipage}[t]{0.5\textwidth}
\leftline{\includegraphics[clip,width=0.8\textwidth]{FigMQ.eps}}
\end{minipage}
\begin{minipage}[t]{0.5\textwidth}
\vspace*{-23ex}

\rightline{\includegraphics[clip,width=1.15\textwidth]{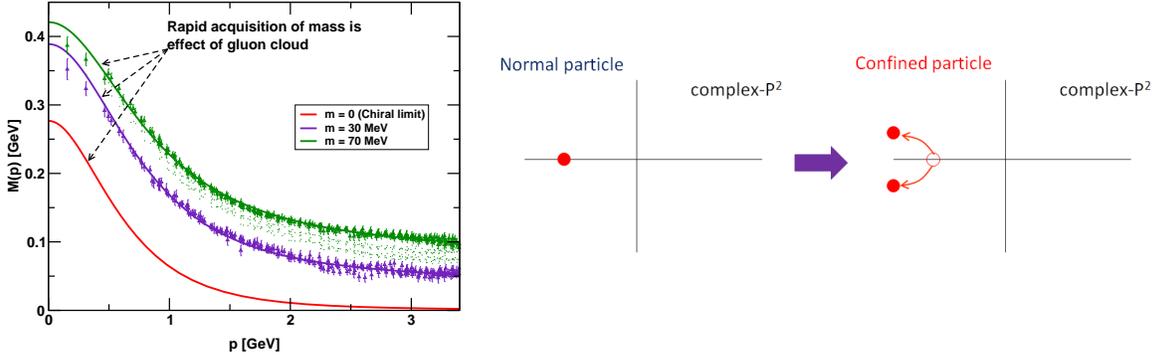}}
\end{minipage}
\end{minipage}
\caption{\label{fig:Fig1}
\underline{Left panel} -- Dressed-quark mass function, $M(p)$: \emph{solid curves} -- DSE results, \protect\cite{Bhagwat:2003vw,Bhagwat:2006tu}, ``data'' -- lattice-QCD simulations \protect\cite{Bowman:2005vx}.  (N.B.\ $m=70\,$MeV is the uppermost curve.  Current-quark mass decreases from top to bottom.)  
The constituent mass arises from a cloud of low-momentum gluons attaching themselves to the current-quark: DCSB is a truly nonperturbative effect that generates a quark mass \emph{from nothing}; namely, it occurs even in the chiral limit, as evidenced by the $m=0$ curve.
\underline{Middle panel} -- An observable particle is associated with a pole at timelike-$P^2$.  This becomes a branch point if, e.g., the particle is dressed by photons.  \underline{Right panel} -- When the dressing interaction is confining, the real-axis mass-pole splits, moving into pairs of complex conjugate poles or branch points.  No mass-shell can be associated with a particle whose propagator exhibits such singularity structure \cite{Krein:1990sf}.
}
\end{figure}

With the introduction of the parton model to describe deep inelastic scattering (DIS), this notion was challenged via an argument \cite{Casher:1974xd} that DCSB can be realised as an intrinsic property of hadrons, instead of through a nontrivial vacuum exterior to the observable degrees of freedom.  Such a view is tenable because the essential ingredient required for dynamical symmetry breaking in a composite system is the existence of a divergent number of constituents and DIS provided evidence for the existence within every hadron of a sea of low-momentum partons.  This perspective has, however, received scant attention.  Instead the introduction of QCD sum rules as a theoretical artifice to estimate nonperturbative strong-interaction matrix elements entrenched the belief that the QCD vacuum is characterised by numerous distinct, spacetime-independent condensates.

Faith in empirical vacuum condensates may be compared with an earlier misguided conviction that the universe was filled with a luminiferous aether.  Recall that physics theories of the late 19th century postulated that, just as water waves must have a medium to move across (water), and audible sound waves require a medium to move through (such as air or water), so also light waves require a medium, the ``luminiferous aether.''  This was apparently unassailable logic until, of course, one of the most famous failed experiments in the history of science \protect\cite{MichelsonMorley}.

Notwithstanding the prevalence of the belief in empirical vacuum condensates, it does lead to problems; e.g., entailing a cosmological constant that is $10^{46}$-times greater than that which is observed \cite{Turner:2001yu,Brodsky:2009zd}.  This unwelcome consequence is partly responsible for reconsideration of the possibility that the so-called vacuum condensates are in fact an intrinsic property of hadrons.  Namely, in a confining theory condensates are not constant, physical mass-scales that fill all spacetime; instead, they are merely mass-dimensioned parameters that serve a practical purpose in some theoretical truncation schemes but otherwise do not have an existence independent of hadrons \cite{Brodsky:2009zd,Brodsky:2010xf,Chang:2011mu,Brodsky:2008be,Glazek:2011vg}.  Confinement is essential to this view and no effort to decry it is meaningful unless the model used in the attempt possesses a veracious expression of this phenomenon.

This factor of $10^{46}$ mismatch is part of what has been called \cite{Turner:2001yu} ``\emph{the greatest embarrassment in theoretical physics}.''  However, it vanishes if one discards the notion that condensates have a physical existence, which is independent of the hadrons that express QCD's asymptotically realisable degrees of freedom \cite{Brodsky:2009zd}; namely, if one accepts that such condensates are merely mass-dimensioned parameters in one or another theoretical truncation scheme.  This appears mandatory in a confining theory \cite{Brodsky:2010xf,Chang:2011mu}, a perspective one may embed in a broader context by considering just what is observable in quantum field theory \cite{Weinberg:1978kz}: ``\ldots \emph{although individual quantum field theories have of course a good deal of content, quantum field theory itself has no content beyond analyticity, unitarity, cluster decomposition and symmetry}.''  If QCD is a confining theory, then cluster decomposition is only realised for colour singlet states \cite{Krein:1990sf} and all observable consequences of the theory, including its ground state, can be expressed via an hadronic basis.  This is quark-hadron duality.

The arguments in Refs.\,\cite{Brodsky:2010xf,Chang:2011mu} follow from two key realisations.  The first is connected with the in-pseudoscalar-meson quark condensate, which is a quantity with an exact expression in QCD; viz. \cite{Maris:1997hd,Maris:1997tm}, $\kappa^\zeta_{P_{f_1 f_2}} = \rho_{P_{f_1 f_2}}^\zeta f_{P_{f_1 f_2}}$, where ($k_\pm = k\pm P/2$)
\begin{eqnarray}
\label{fpigen}
i f_{P_{f_1 f_2}} K_\mu = \langle 0 | \bar q_{f_2} \gamma_5 \gamma_\mu q_{f_1} |P \rangle &=&  Z_2(\zeta,\Lambda)\; {\rm tr}_{\rm CD}
\int_k^\Lambda i\gamma_5\gamma_\mu S_{f_1}(k_+) \Gamma_{P_{f_1 f_2}}(k;K) S_{f_2}(k_-)\,, \\
i\rho_{P_{f_1 f_2}}^\zeta = -\langle 0 | \bar f_2 i\gamma_5 f_1 |P \rangle
&=& Z_4(\zeta,\Lambda)\; {\rm tr}_{\rm CD}
\int_k^\Lambda \gamma_5 S_{f_1}(k_+) \Gamma_{P_{f_1 f_2}}(k;K) S_{f_2}(k_-) \,,\label{rhogen}\\
\label{GMORP}
f_{P_{f_1 f_2}}^2 m_{P_{f_1 f_2}}^2 &=& [m_{f_1}^\zeta +m_{f_2}^\zeta]\, \kappa^\zeta_{P_{f_1 f_2}}.
%
%
\end{eqnarray}
Here $\int_k^\Lambda$ is a Poincar\'e-invariant regularization of the integral, with $\Lambda$ the ultraviolet regularization mass-scale, $Z_{2,4}$ are renormalisation constants, with $\zeta$ the renormalisation point, $\Gamma_{P_{f_1 f_2}}$ is the meson's Bethe-Salpeter amplitude, and $S_{f_1,f_2}$ are the component dressed-quark propagators, with $m_{f_1,f_2}$ the associated current-quark masses.  Equation~(\ref{fpigen}) describes the pseudoscalar meson's leptonic decay constant; i.e., the pseudovector projection of the meson's Bethe-Salpeter wave-function onto the origin in configuration space; Eq.\,(\ref{rhogen}) describes its pseudsocalar analogue; and $m_{P_{f_1 f_2}}$ is the meson's mass.  The initial step in extending the concept was a proof that the in-pseudoscalar-meson quark condensate, $\kappa^\zeta_{P_{f_1 f_2}}$, can rigorously be represented through the pseudoscalar-meson's scalar form factor at zero momentum transfer, $Q^2=0$; viz., ($m^\zeta_{f_1 f_2}=m^\zeta_{f_1}+ m^\zeta_{f_2}$)
\begin{equation}
{\cal S}_{P_{f_1 f_2}} := -\langle P_{f_1 f_2}|\mbox{\small $\frac{1}{2}$} (\bar q_{f_1} q_{f_2}+\bar q_{f_1} q_{f_2}) | P_{f_1 f_2} \rangle
=\frac{\partial m^2_{P_{f_1 f_2}}}{\partial m_{f_1 f_2}^\zeta}
=
\frac{ \kappa^\zeta_{P_{f_1 f_2}} }{f^2_{P_{f_1 f_2}}} + m_{f_1 f_2}^\zeta
\frac{\partial}{\partial m_{f_1 f_2}^\zeta}  \left[ \frac{ \kappa^\zeta_{P_{f_1 f_2}} }{f^2_{P_{f_1 f_2}}}\right].
\end{equation}

The second step employs a mass formula for scalar mesons, exact in QCD; viz. \cite{Chang:2011mu},
\begin{eqnarray}
\label{gmorS}
f_{S_{f_1 f_2}} m_{S_{f_1 f_2}}^2 &=& -[m_{f_1}^\zeta - m_{f_2}^\zeta] \, \rho_{S_{12}}^\zeta,\\
f_{S_{f_1 f_2}} K_\mu
& = & Z_2\, {\rm tr}_{\rm CD}\!\!\!
\int_k^\Lambda i\gamma_\mu S_{f_1}(k_+) \Gamma_{S_{f_1 f_2}}(k;K) S_{f_2}(k_-)\,, \rule{2em}{0ex}
\label{fsigmagen} \\
\rho^\zeta_{S_{f_1 f_2}}
& = & - Z_4\, {\rm tr}_{\rm CD}\!\!\!
\int_k^\Lambda S_{f_1}(k_+) \Gamma_{S_{f_1 f_2}}(k;K) S_{f_2}(k_-) , \rule{2em}{0ex}\label{rhosigmagen}
\end{eqnarray}
which was used to prove that the in-scalar-meson quark condensate is, analogously and rigorously, connected with the scalar-meson's scalar form factor at $Q^2=0$.  Moreover the following limiting cases were also established:
\begin{equation}
\label{HQlimits}
f_{P,S}^2 \, {\cal S}^\zeta_{P,S} \stackrel{\rm chiral\,limit}{=} \kappa^0 = -\langle \bar q q \rangle^0
\; \mbox{and}\;
f_{P,S}^2 \, {\cal S}^\zeta_{P,S} \stackrel{\rm heavy\,quark(s)}{=} 2 \kappa^\zeta_{P,S},
\end{equation}
where $\langle \bar q q \rangle^0$ is precisely the quantity that is widely known as the \emph{vacuum quark condensate} through any of its definitions \cite{Langfeld:2003ye}.

With appeal then to demonstrable results of heavy-quark symmetry in QCD, it was argued that the $Q^2=0$ values of vector- and pseudovector-meson scalar form factors also determine the in-hadron condensates in these cases, and that this expression for the concept of in-hadron quark condensates is readily extended to the case of baryons.  Thus, through the $Q^2=0$ value of any hadron's scalar form factor, one can extract the magnitude of a quark condensate in that hadron which is a reasonable and realistic measure of dynamical chiral symmetry breaking.


Note that in the presence of confinement it is impossible to write a valid nonperturbative definition of a single quark or gluon annihilation operator; and therefore impossible to rigorously define a second quantised ground state for QCD upon a foundation of gluon and quark (quasiparticle) operators.   To do so would be to answer the question: What is the state that is annihilated by an operator which is -- as appears at present -- unknowable?  However, with the assumptions that confinement is absolute and that it entails quark-hadron duality, the question changes completely.  Then the nonperturbative Hamiltonian of observable phenomena in QCD is diagonalised by colour-singlet states alone.  The ground state of this strong-interaction Hamiltonian is the state with zero hadrons.  One may picture the creation and annihilation operators for such states as rigorously defined via smeared sources on a spacetime lattice.  The ground-state is defined with reference to such operators, employing, e.g., the Gell-Mann - Low theorem \cite{GellMann:1951rw}, which is applicable in this case because there are well-defined asymptotic states and associated annihilation and creation operators.

In learning that the so-called vacuum quark condensate is actually the chiral-limit value of an in-pion property, some respond as follows.
The electromagnetic radius of any hadron, $r_H^{\rm em}$, which couples to pseudoscalar mesons must diverge in the chiral limit.  This long-known effect arises because the propagation of \emph{massless} on-shell colour-singlet pseudoscalar mesons is undamped \cite{Alkofer:1993gu}.
Therefore, does not each pion grow to fill the universe; so that, in this limit, the in-pion condensate reproduces the conventional paradigm?

Confinement, again, vitiates this objection.  Both DSE- and lattice-QCD indicate that confinement entails dynamical mass generation for both gluons and quarks, Fig.\,\ref{fig:Fig1}.  The dynamical gluon and quark masses remain large in the limit of vanishing current-quark mass.  In fact, the dynamical masses are almost independent of current-quark mass in the neighbourhood of the chiral limit.  It follows that for any hadron the quark-gluon containment-radius does not diverge in the chiral limit.  Instead, it is almost insensitive to the magnitude of the current-quark mass because the dynamical masses of the hadron's constituents are frozen at large values; viz., $2 - 3\,\Lambda_{\rm QCD}$.  These considerations show that the divergence of $r_\pi^{\rm em}$ does not correspond to expansion of a condensate from within the pion but rather to the copious production and subsequent propagation of composite pions, each of which contains a condensate whose value is essentially unchanged from its nonzero current-quark mass value within a containment-domain whose size is similarly unaffected.

There is more to be said in connection with the definition and consequences of a chiral limit.  Plainly, the existence of strongly-interacting massless composites would have an enormous impact on the evolution of the universe; and it is naive to imagine that one can simply set the renormalisation-point-invariant current-quark masses to zero and consider a circumscribed range of manageable consequences whilst ignoring the wider implications for hadrons, the Standard Model and beyond.  For example, with all else held constant, Big Bang Nucleosynthesis is very sensitive to the value of the pion-mass \cite{Flambaum:2007mj}.  We are fortunate that the absence of quarks with zero current-quark mass has produced a universe in which we exist so that we may carefully ponder the alternative.

With quark condensates being an intrinsic property of hadrons, one arrives at a new paradigm, as observed in the popular science press \cite{Courtland:2010zz}: ``\emph{EMPTY space may really be empty.  Though quantum theory suggests that a vacuum should be fizzing with particle activity, it turns out that this paradoxical picture of nothingness may not be needed.  A calmer view of the vacuum would also help resolve a nagging inconsistency with dark energy, the elusive force thought to be speeding up the expansion of the universe}.''  In connection with the cosmological constant, putting QCD condensates back into hadrons reduces the mismatch between experiment and theory by a factor of $10^{46}$.  If technicolour-like theories are the correct scheme for extending the Standard Model \cite{Andersen:2011yj}, then the impact of the notion of in-hadron condensates is far greater still.

\section{Dressed-quark Anomalous Moments}
\label{sec:ACM}
The appearance and behaviour of $M(p)$ in Fig.\,\ref{fig:Fig1} are essentially quantum field theoretic effects, unrealisable in quantum mechanics.  The running mass connects the infrared and ultraviolet regimes of the theory, and establishes that the constituent-quark and current-quark masses are simply two connected points on a single curve separated by a large momentum interval.  QCD's dressed-quark behaves as a constituent-quark, a current-quark, or something in between, depending on the momentum of the probe which explores the bound-state containing the dressed-quark.  These remarks should make clear that QCD's dressed-quarks are not simply Dirac particles.

Since the two-point functions of elementary excitations are strongly modified in the infrared, then the same is generally true for three-point functions; i.e., the vertices.  The bare vertex will thus be a poor approximation to the complete result unless there are extenuating circumstances.  This is readily made apparent, for with a dressed-quark propagator characterised by a momentum-dependent mass-function, one finds immediately that the Ward-Takahashi identity is breached; viz.,
\begin{equation}
P_\mu i \gamma_\mu \neq S^{-1}(k+P/2) - S^{-1}(k-P/2)\,,
\end{equation}
and the violation is significant whenever and wherever the mass function in Fig.\,\ref{fig:Fig1} is large.  The most important feature of the perturbative or bare vertex is that it cannot cause spin-flip transitions; namely, it is an helicity conserving interaction.  However, one must expect that DCSB introduces nonperturbatively generated structures that very strongly break helicity conservation.  These contributions will be large when the dressed-quark mass-function is large.  Conversely, they will vanish in the ultraviolet; i.e., on the perturbative domain.  The exact form of the vertex contributions is still the subject of study but their existence is model-independent.

A spectacular consequence is highlighted by considering dressed-quark anomalous magnetic moments.  In QCD the analogue of Schwinger's one-loop calculation can be performed to find an anomalous \emph{chromo}-magnetic moment (ACM) for the quark.  There are two diagrams: one similar in form to that in QED; and another owing to the gluon self-interaction.  One reads from Ref.\,\cite{Davydychev:2000rt} that the perturbative result vanishes in the chiral limit.  This is consonant with helicity conservation in perturbative massless-QCD.  However, Fig.\,\ref{fig:Fig1} demonstrates that chiral symmetry is dynamically broken strongly in QCD and one must therefore ask whether this affects the ACM.

Of course, it does, and it is now known that the effect is modulated by the strong momentum dependence of the dressed-quark mass-function \cite{Chang:2010hb}.  In fact, dressed-quarks possess a large, dynamically-generated ACM, which produces an equally large ano\-malous electromagnetic moment that has a material impact on nucleon magnetic and transition form factors \cite{Chang:2011tx,Wilson:2011rj}.  Furthermore, given the magnitude of the muon ``$g_\mu-2$ anomaly'' and its assumed importance as an harbinger of physics beyond the Standard Model \cite{Jegerlehner:2009ry}, it might also be worthwhile to make a quantitative estimate of the contribution to $g_\mu-2$ from the quark's DCSB-induced anomalous moments following, e.g., the computational pattern indicated in Ref.\,\cite{Goecke:2011pe}.

We now illustrate an interesting possibility, which has recently been identified; viz., that the dressed-quark's ACM is self-limiting.  Consider the gap equation obtained from a symmetry-preserving regularisation of a vector-vector contact interaction \cite{Wilson:2011rj} with a quark-gluon vertex of the form $\Gamma_\nu^a = \Gamma_\nu \lambda^a/2$ [$k=p_f-p_i$, $\ell = (p_f+p_i)/2$],
\begin{eqnarray}
\nonumber
\Gamma_\mu(p_f,p_i;k)&=& \gamma_\mu \, A
+ \sigma_{\mu\nu} k_\nu \tau_2
+ \sigma_{\mu\nu} \ell_\nu \,\ell\cdot k\, \tau_3
+ [\ell_\mu \gamma\cdot  k + i \gamma_\mu \sigma_{\nu\rho}\ell_\nu k_\rho]  \tau_4 \\
& &
- i \ell_\mu \tau_5 +\, \ell_\mu \gamma\cdot k \, \ell \cdot  k\, \tau_6 - \ell_\mu \gamma\cdot \ell \, \tau_7  + \ell_\mu \sigma_{\nu\rho} \ell_\nu k_\rho \tau_8\,.
\end{eqnarray}
The dressed-quark propagator has the form $S(p) = 1/(i A \gamma \cdot p + B)=(1/A)/(i\gamma \cdot p + M)$, where
\begin{subequations}
\label{gapGamma}
\begin{eqnarray}
    A &=& 1 + \frac{16 \pi \alpha_{\rm IR}}{3 m_G^2}
    \int\frac{d^4q}{(2\pi)^4}  \frac{3}{2} \frac{M}{A}
    \frac{1}{q^2  +M^2} \,\mu_Q \,(\hat \tau_5 -2)\,, \\
    B &=& m + \frac{16 \pi \alpha_{\rm IR}}{3 m_G^2}
    \int\frac{d^4q}{(2\pi)^4} \frac{M}{A} \, \frac{1}{q^2+M^2} \, ( 4 A - 7 M\, \mu_Q) \,,
\end{eqnarray}
\end{subequations}
with $m_G=0.8\,$GeV, a mass-scale typical of gluons \cite{Bowman:2004jm,Aguilar:2009nf,Boucaud:2011ug,Qin:2011dd}; and $\alpha_{\rm IR}$ a parameter characterising the infrared strength of the interaction.  A context for the value of $\alpha_{\rm IR}$ is easily provided.
In rainbow-ladder truncation with nonperturbatively-massless gauge bosons, the coupling below which DCSB breaking is impossible via the gap equations in QED and QCD is $\alpha^c/\pi \approx 1/3$ \cite{Bloch:2002eq,Bashir:2011dp},
whilst with a symmetry-preserving regularisation of a contact-interaction, $\alpha^c_{\rm IR}/\pi \approx 0.4$ and a description of hadron phenomena requires $\alpha_{\rm IR}/\pi \approx 1$ \cite{Wilson:2011rj}.  The simplicity of Eqs.\,\eqref{gapGamma} was attained as follows: we imposed two constraints from perturbative QCD \cite{Davydychev:2000rt}, $\tau_2 + M \tau_4 = -\mu_Q + \tau_5/2 + M \tau_7/2$, $\mu_Q>0$; wrote $M \tau_7 = \tau_5 = (2+\hat\tau_5) \mu_Q$ in order to suppress a parameter; and used $p_f^2 = M^2 = p_i^2$, identifying the focus of integration strength, in order to simplify the integrands.

One may now read that if $\tau_5$, $\tau_7$ are large enough; viz., $\hat \tau_5 \geq 2$, then $A\geq 1$, in which case a $B>0$ solution is possible so long as $M \mu_Q < (4A/7)$.  Hence, there is an internally consistent upper bound on the dressed-quark ACM.
Furthermore, including an ACM in the gap equation produces the same sort of attraction as does introducing resonant contributions via the vertex -- the so-called meson-loop effects \cite{Cloet:2008fw}.  One should thus expect that their strength, too, is self-limiting.
Finally, this simple analysis confirms a long-known result that dressing the vertex; i.e., improving sensibly upon the rainbow-ladder truncation, decreases the $p^2=0$ value of the mass-function: strength is shifted out to $p\gtrsim 2 \Lambda_{\rm QCD}$ \cite{Bhagwat:2004hn}.
Therefore one should not expect vertex dressing to uniformly boost the magnitude of the dressed-quark mass function nor to materially reduce the critical coupling below which  DCSB is impossible \cite{Bashir:2011dp}.

\section{Deep Inelastic Scattering}
The past forty years have seen a tremendous effort to deduce the parton distribution
functions (PDFs) of the most accessible hadrons -- the proton, neutron and pion.  There are many reasons for this long sustained and thriving interest \cite{Holt:2010vj} but in large part it is motivated by the suspected process-independence of the usual parton distribution functions and hence an ability to unify many hadronic processes through their computation.  In connection with uncovering the essence of the strong interaction, the behaviour of the valence-quark distribution functions at large Bjorken-$x$ is most relevant.
Furthermore, an accurate determination of the behavior of distribution functions in the valence region is also important to high-energy physics.  Particle discovery experiments and Standard Model tests with colliders are only possible if the QCD background is completely understood.  QCD evolution, apparent in the so-called scaling violations by parton distribution functions,\footnote{DGLAP evolution is described in Sec.IID of Ref.\,\cite{Holt:2010vj}.  The evolution equations are derived in perturbative QCD and determine the rate of change of parton densities when the energy-scale chosen for their definition is varied.}
entails that with increasing center-of-mass energy, the support at large-$x$ in the distributions evolves to small-$x$ and thereby contributes materially to the collider background.

Owing to the dichotomous nature of Goldstone bosons, understanding the valence-quark distribution functions in the pion and kaon is of great importance.  Moreover, given the large value of the ratio of $s$-to-$u$ current-quark masses, a comparison between the pion and kaon structure functions offers the chance to chart effects of explicit chiral symmetry breaking on the structure of would-be Goldstone modes.  There is also the prediction \cite{Ezawa:1974wm,Farrar:1975yb} that a theory in which the quarks interact via $1/k^2$-vector-boson exchange will produce valence-quark distribution functions for which
\begin{equation}
\label{pQCDuvx}
q_{\rm v}(x) \propto (1-x)^{2+\gamma} \,,\; x\gtrsim 0.85\,,
\end{equation}
where $\gamma\gtrsim 0$ is an anomalous dimension that grows with increasing momentum transfer.  (See Sec.VI.B.3 of Ref.\,\cite{Holt:2010vj} for a detailed discussion.)

\begin{figure}[t]
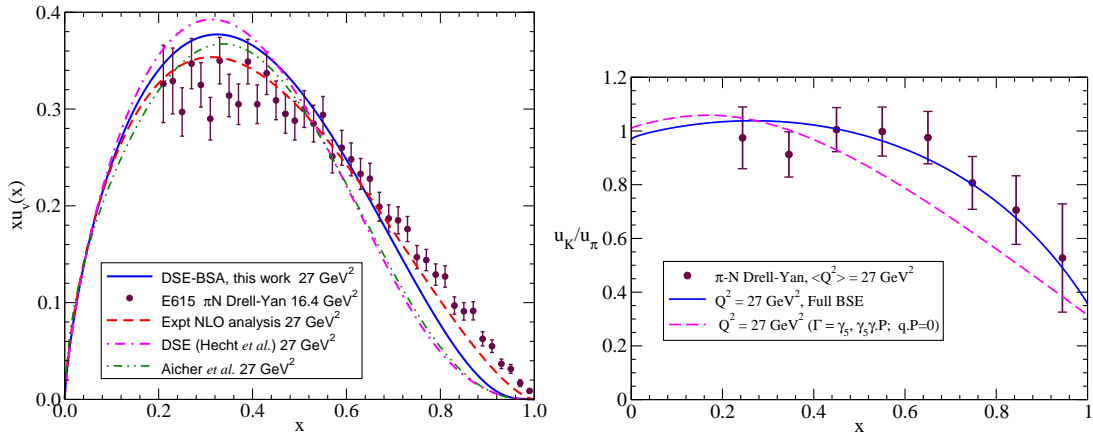

\begin{minipage}[t]{0.95\textwidth}
\begin{minipage}[t]{0.5\textwidth}
\leftline{\includegraphics[clip,width=1.0\textwidth]{Fig2PDFCJP.eps}}
\end{minipage}
\begin{minipage}[t]{0.5\textwidth}
\rightline{\includegraphics[clip,width=1.0\textwidth]{Fig4PDFCJP.eps}}
\end{minipage}
\end{minipage}

\caption{
\underline{Left panel}. Pion valence quark distribution function.  \emph{Solid curve} -- full DSE calculation \protect\cite{Nguyen:2011jy}; \emph{dot-dashed curve} -- semi-phenomenological DSE-based calculation in Ref.\,\protect\cite{Hecht:2000xa}; \emph{filled circles} -- experimental data from Ref.\,\protect\cite{Conway:1989fs};
\emph{dashed curve} -- NLO reanalysis of the experimental data \protect\cite{Wijesooriya:2005ir};
and \emph{dot-dot-dashed curve} -- NLO reanalysis of experimental data with inclusion of soft-gluon resummation \protect\cite{Aicher:2010cb}.
\underline{Right panel}.  DSE prediction for the ratio of $u$-quark distributions in the kaon and pion \protect\cite{Holt:2010vj,Nguyen:2011jy}.  The full Bethe-Salpeter amplitude produces the \emph{solid} curve; a reduced BSE vertex produces the \emph{dashed} curve.  The reduced amplitude retains only the invariants and amplitudes  involving pseudoscalar and axial vector Dirac matrices, and ignores dependence on the variable $q\cdot P$.  These are part of the reductions that occur in a pointlike treatment of pseudoscalar mesons.  The experimental data is from \protect\cite{Badier:1980jq,Badier:1983mj}
\label{fig:pi_DSE}}
\end{figure}

Since pseudoscalar meson targets are difficult to produce, experimental knowledge of the parton structure of the pion and kaon arises primarily from pionic or kaonic Drell-Yan processes; and such an experiment \cite{Conway:1989fs} produced very disturbing results: a valence-quark distribution function in striking conflict with QCD.  Instead of Eq.\,\eqref{pQCDuvx}, the data indicated $u_V(x)\sim(1-x)^1$.  The exponent ``1'' is that predicted by a theory without gluons; i.e., constituent-quark-like or contact-interaction models.  Many authors employed such models to reproduce the data and argued therefrom that inferences from QCD were wrong.  This critical disagreement was re-emphasized by nonperturbative DSE predictions \cite{Nguyen:2011jy,Hecht:2000xa} which confirmed the exponent ``2'' and demonstrated its connection with the dressed-quark mass-function that is the smoking gun for dynamical chiral symmetry breaking.  Not until very recently was a resolution of the conflict between data and well-constrained theory presented \cite{Aicher:2010cb}: it confirms the DSE predictions; and thereby emphasises the predictive power and strength of using a single internally-consistent, well-constrained framework to correlate and unify the description of hadron observables.

With Ref.\,\cite{Nguyen:2011jy} a significant milestone was achieved: the first unification of the computation of distribution functions that arise in analyses of deep inelastic scattering with that of numerous other properties of pseudoscalar mesons, including meson-meson scattering \cite{Bicudo:2001jq} and the successful prediction of electromagnetic elastic and transition form factors \cite{Maris:2000sk,Roberts:2010rn}.  The results confirmed the large-$x$ behavior of distribution functions predicted by the QCD parton model; provide a good account of the $\pi N$ Drell-Yan data for $u_V^\pi(x$); and a parameter-free prediction for the ratio $u_V^K(x)/u_V^\pi(x)$ that agrees with extant data, showing a strong environment-dependence of the $u$-quark distribution (right panel in Fig.\,\ref{fig:pi_DSE}).  The new Drell-Yan experiment running at FNAL is capable of validating this comparison, as is the COMPASS~II experiment at CERN.  Such an experiment should be done so that complete understanding of QCD's Goldstone modes can be claimed.

These successes should be contrasted with attempts to compute parton distribution functions in constituent-quark-like models.  Such quantum mechanical models cannot incorporate the mo\-men\-tum-dependent dressed-quark and dressed-gluon mass-functions; they have no connection with quantum field theory and therefore not with QCD; and they are not ``symmetry-preserving'' and hence cannot veraciously connect meson and baryon properties.
Amongst their defects, such models typically produce very hard distributions.  In order then to make contact with the real world's soft distributions, their adherents are forced to choose an unrealistically small value for the so-called hadronic scale, $Q_0$, from which DGLAP evolution must begin; viz., $Q_0 \sim 1.25 \,\Lambda_{\rm QCD}$.  There are no sound reasons to support an assertion that perturbative evolution can be used at such low scales.  Indeed, it is plain from inspection of a plot of the dressed-quark mass function, Fig.\,\ref{fig:Fig1}, that the domain $Q_0<2\,\Lambda_{\rm QCD}$ is \emph{essentially} nonperturbative because thereupon $M(p^2)$ exhibits the inflexion point that is characteristic of confinement.  Thus any use of DGLAP from such low mass-scales is misguided.  As made plain by the experience with $u_V^\pi(x)$, it will only express model artefacts and yield no reliable information about QCD.

The hardness of the distributions in such models can be traced to the complete lack of gluons, which is represented via enforcement of what may be called the ``constituent-quark momentum sum rule;'' viz,
\begin{equation}
\langle x  \rangle_{Q_0}
= \int_0^1 dx \, x \sum_{i=1}^N \,q_V(x;Q_0) = 1
\end{equation}
where $N$ is the number of valence-quarks in the hadron.  As elucidated in Sec.VI.B.3 of Ref.\,\cite{Holt:2010vj}, in QCD there is no scale at which the valence quarks can carry all of a hadron's momentum.  This is readily explained.  All hadrons are bound by gluons, which are invisible to the electromagnetic probe.  Thus, some fraction of the hadron's momentum is carried by gluons at all resolving scales unless the hadron is a point particle.  Indeed, it is a simple algebraic exercise to demonstrate that the only non-increasing, convex function which can produce $\langle x^0\rangle =N$ and $\langle x \rangle = 1$, is the distribution $q_V(x)=1$, which is uniquely connected with a pointlike hadron; viz., a hadron whose bound-state amplitude is momentum-independent.  The real consequence of the momentum sum-rule in QCD is that at any resolving scale, $Q$,
\begin{equation}
\int_0^1 dx \, x \sum_{i=1}^N \,q_V(x;Q) < 1 \,.
\end{equation}

As emphasised above, PDFs can be an excellent probe of the running coupling but only if the model input is rigorously connected with QCD, as with the DSE prediction of the pion's valence-quark distribution function \cite{Hecht:2000xa}, which was verified by the new NLO analysis of extant data \cite{Aicher:2010cb}.

\section{Grand Unification}
Owing to the importance of DCSB, full capitalisation on the results of forthcoming experimental programmes is only possible if the properties of meson and baryon ground- and excited-states can be correlated within a single, symmetry-preserving framework, where symmetry-preserving means that all relevant Ward-Takahashi identities are satisfied.  Constituent-quark-like models, which cannot incorporate the momentum-dependent dressed-quark mass-function, fail this test.

An alternative is being pursued within quantum field theory via the Faddeev equation.  This analogue of the Bethe-Salpeter equation sums all possible interactions that can occur between three dressed-quarks.  A simplified equation \cite{Cahill:1988dx} is founded on the observation that an interaction which describes color-singlet mesons also generates nonpointlike quark-quark (diquark) correlations in the color-antitriplet channel \cite{Cahill:1987qr}.  The dominant correlations for ground state octet and decuplet baryons are scalar and axial-vector diquarks because, e.g., the associated mass-scales are smaller than the baryons' masses and their parity matches that of these baryons.  (Recent studies confirm the fidelity of the nonpointlike diquark approximation within the nucleon three-body problem; e.g. Ref.\,\protect\cite{Eichmann:2011vu}.) On the other hand, pseudoscalar and vector diquarks dominate in the parity-partners of those ground states \cite{Roberts:2011cf}.  This approach treats mesons and baryons on the same footing and, in particular, enables the impact of DCSB to be expressed in the prediction of baryon properties.

Building on lessons from meson studies \cite{Chang:2011vu}, a unified spectrum of $u,d$-quark hadrons was obtained using a symmetry-preserving regularisation of a vector$\,\times\,$vector contact interaction \cite{Roberts:2011cf}.  That study correlates the masses of meson and baryon ground- and excited-states within a single framework.  In comparison with relevant quantities, the computation produces $\overline{\mbox{rms}}$=13\%, where $\overline{\mbox{rms}}$ is the root-mean-square-relative-error$/$degree-of freedom.  The predictions uniformly overestimate the PDG values of meson and baryon masses \cite{Nakamura:2010zzi}.  Given that the employed truncation deliberately omitted meson-cloud effects in the Faddeev kernel, this is a good outcome, since inclusion of such contributions acts to reduce the computed masses by just the required amount.

\begin{figure}[t]
\vspace*{5ex}
\begin{minipage}[t]{0.95\textwidth}
\begin{minipage}[t]{0.5\textwidth}
\leftline{\includegraphics[clip,width=1.0\textwidth]{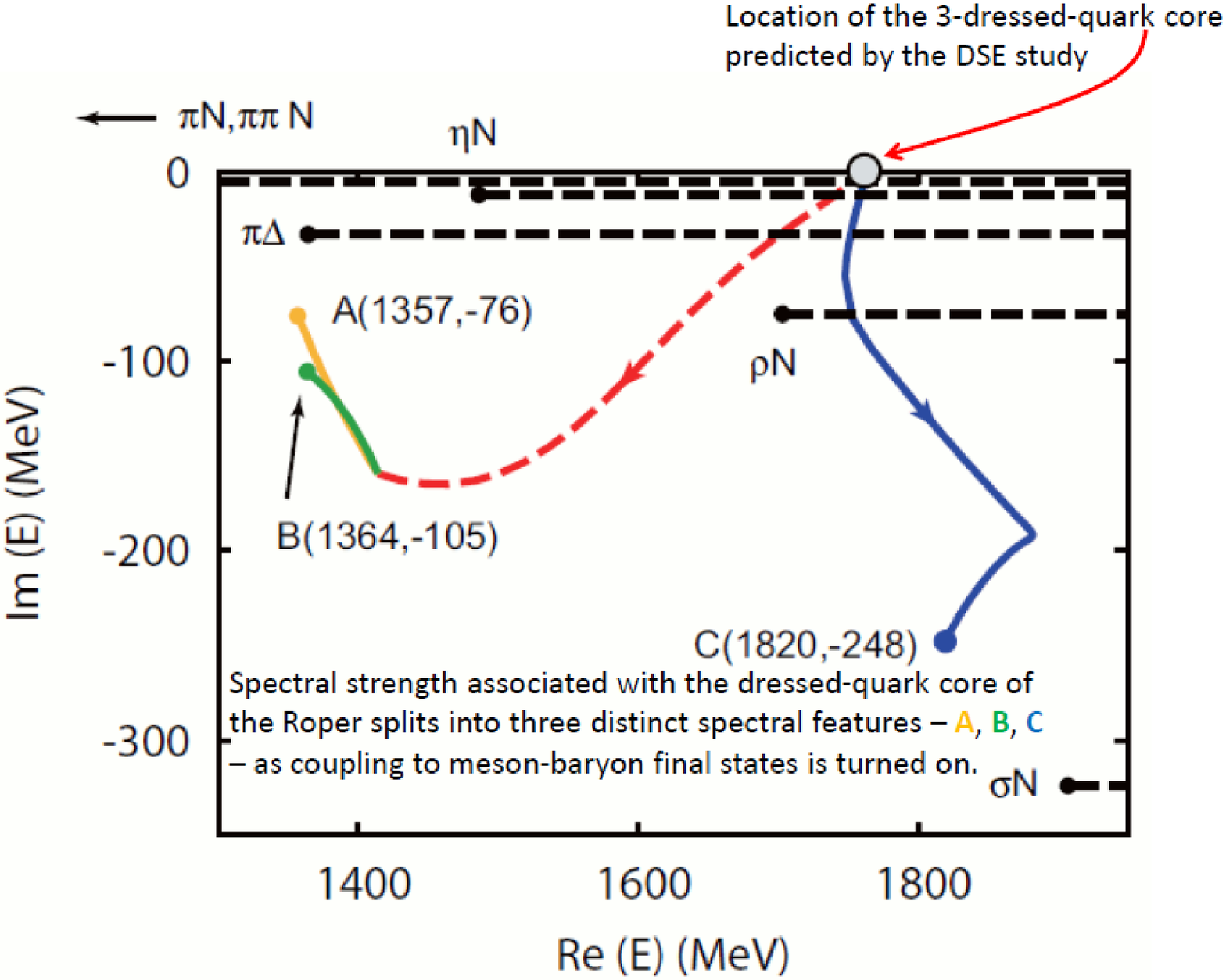}}
\end{minipage}
\begin{minipage}[t]{0.5\textwidth}
\vspace*{-42ex}

\rightline{\includegraphics[clip,width=1.0\textwidth]{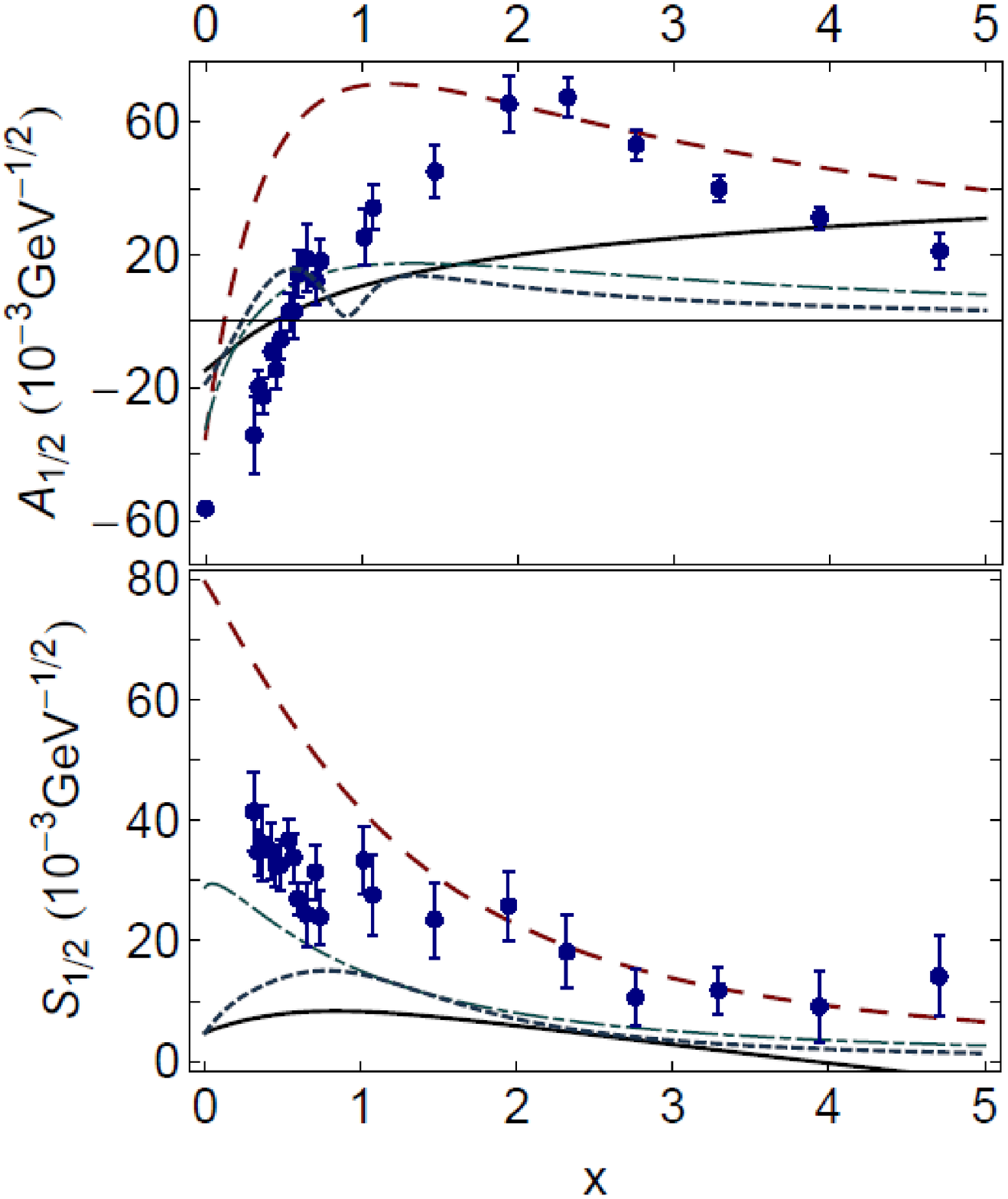}}
\end{minipage}
\end{minipage}

\caption{\label{fig:ebac}
\underline{Left panel}.  The Excited Baryon Analysis Center (EBAC) examined the $P_{11}$-channel and found that the two poles associated with the Roper resonance and the next higher resonance were all associated with the same dressed-quark-core state.  Coupling to the continuum of meson-baryon final states induces multiple observed resonances from the same bare state.  In EBAC's analysis, all PDG-identified resonances were found to consist of a core state plus meson-baryon components.
\underline{Right panel}.
Helicity amplitudes for $\gamma^\ast p \to P_{11}(1440)$, with $x=Q^2/m_N^2$, $m_N$ is the nucleon mass.
Solid curves -- computed using the treatment of a contact interaction described in Ref.\,\protect\cite{Wilson:2011rj}, including the dressed-quark anomalous magnetic moment (Sec.\,\protect\ref{sec:ACM});
dashed curves -- the light-front constituent quark model results from Ref.\,\protect\cite{Aznauryan:2007ja};
long-dash-dot curves -- the light-front constituent quark model results from Ref.\,\protect\cite{Cardarelli:1996vn};
short-dashed curves -- smooth fit in Ref.\,\protect\cite{Wilson:2011rj} to the bare form factors inferred in Ref.\,\protect\cite{Suzuki:2010yn,JuliaDiaz:2009ww};
and data -- Refs.\,\protect\cite{Dugger:2009pn,Aznauryan:2009mx,Aznauryan:2011td}.
}

\end{figure}

Following this reasoning, a striking result is agreement between the DSE-computed baryon masses \cite{Wilson:2011rj,Roberts:2011cf} and the bare masses employed in modern coupled-channels models of pion-nucleon reactions \cite{Suzuki:2009nj,Gasparyan:2003fp}.  The Roper resonance is very interesting.  The DSE study \cite{Wilson:2011rj} produces an excitation of the nucleon at $1.72\pm0.07\,$GeV.  This state is predominantly a radial excitation of the quark-diquark system, with both the scalar- and axial-vector diquark correlations in their ground state.  Its predicted mass lies precisely at the value determined in the analysis of Ref.\,\cite{Suzuki:2009nj}.  This is significant because for almost 50 years the ``Roper resonance'' has defied understanding.  Discovered in 1963, it appears to be an exact copy of the proton except that its mass is 50\% greater.  The mass was the problem: hitherto it could not be explained by any symmetry-preserving QCD-based tool.  That has now changed.  Combined, see Fig.\,\ref{fig:ebac}, Refs.\,\cite{Wilson:2011rj,Roberts:2011cf,Suzuki:2009nj} demonstrate that the Roper resonance is indeed the proton's first radial excitation, and that its mass is far lighter than normal for such an excitation because the Roper obscures its dressed-quark-core with a dense cloud of pions and other mesons.  Such feedback between QCD-based theory and reaction models is critical now and for the foreseeable future, especially since analyses of experimental data on nucleon-resonance electrocouplings suggest strongly that this structure is typical; i.e., most low-lying $N^\ast$-states can best be understood as an internal quark-core dressed additionally by a meson cloud \cite{Aznauryan:2011td,Gothe:2011up}.

With masses and Faddeev amplitudes in hand, it is possible to compute baryon electromagnetic elastic and transition form factors.  The most extensive studies are presented in Refs.\,\cite{Wilson:2011rj,Eichmann:2011vu,Cloet:2008re}, which unify the computation of meson and nucleon form factors, and also their valence-quark distribution functions.  These studies demonstrate that elastic scattering and resonance electroproduction experiments probe the evolution of the strong interaction's running masses and coupling to infrared momenta.  Clear signals are found, for example \cite{Chang:2011tx,Wilson:2011rj}: in the existence, and location if so, of a zero in the ratio of nucleon Sachs form factors, which are strongly influenced by the running of the dressed-quark mass; and in the impact of the dressed-quark anomalous electromagnetic moment on the proton's magnetic form factor.  They also emphasise that on the domain $0 < Q^2\lesssim 2\,$GeV$^2$, meson-cloud effects are important in making a realistic comparison between experiment and hadron structure calculations.  The computed $\gamma^\ast p \to P_{11}(1440)$ helicity amplitudes \cite{Wilson:2011rj} are similar to the bare amplitudes inferred via coupled-channels analyses of the electroproduction process (Fig.\,\ref{fig:ebac}, right panel).  This supports a view that extant hadron structure calculations, which typically omit meson-cloud effects, should only be directly compared with the bare-masses, -couplings, etc., determined via coupled-channels analyses \cite{Suzuki:2010yn,JuliaDiaz:2009ww}.

It is also worth reviewing a connection between the $Q^2=0$ values of elastic form factors and the Bjorken-$x=1$ values of the DIS structure functions, $F_2^{n,p}(x)$.  First recall that the $x=1$ value of a structure function is invariant under the evolution equations \cite{Holt:2010vj}.  Hence the value of
\begin{equation}
\label{dvuv1}
\left. \frac{d_v(x)}{u_v(x)}\right|_{x\to 1}\rule{-0.5em}{0ex}, \;\mbox{where} \rule{1em}{0ex}
\frac{d_v(x)}{u_v(x)} =
\frac{4 \frac{F_2^n(x)}{F_2^p(x)} - 1}{4- \frac{F_2^n(x)}{F_2^p(x)}},
\end{equation}
is a scale-invariant feature of QCD and a discriminator between models.  Next, when Bjorken-$x$ is unity, then $Q^2+2P\cdot Q=0$; i.e., one is dealing with elastic scattering.  Therefore, in the neighbourhood of $x=1$ the structure functions are determined by the target's elastic form factors.
Equation~\eqref{dvuv1} expresses the relative probability of finding a $d$-quark carrying all the proton's light-front momentum compared with that of a $u$-quark doing the same or, equally, owing to invariance under evolution, the relative probability that a $Q^2=0$ probe either scatters from a $d$-quark or a $u$-quark; viz.,
\begin{equation}
\label{dvuvF1}
\left. \frac{d_v(x)}{u_v(x)}\right|_{x\to 1} = \frac{P_{1}^{p,d}}{P_{1}^{p,u}}.
\end{equation}
%

In constituent-quark models with $SU(6)$-symmetric spin-flavour wave-functions the right-hand-side of Eq.\,\eqref{dvuvF1} is $1/2$ because there is nothing to distinguish between the wave-functions of $u$- and $d$-quarks, and the proton is constituted from $u$-quarks and one $d$-quark.  On the other hand, when a Poincar\'e-covariant Faddeev equation is employed to describe the nucleon,
\begin{equation}
\label{dvuvF1result}
\frac{P_{1}^{p,d}}{P_{1}^{p,u}} =
\frac{\frac{2}{3} P_1^{p,a} + \frac{1}{3} P_1^{p,m}}
{P_1^{p,s}+\frac{1}{3} P_1^{p,a} + \frac{2}{3} P_1^{p,m}},
\end{equation}
where we have used the notation of Ref.\,\cite{Cloet:2008re}.  Namely,
$P_1^{p,s}=F_{1p}^s(Q^2=0)$ is the contribution to the proton's charge arising from diagrams with a scalar diquark component in both the initial and final state.  The diquark-photon interaction is far softer than the quark-photon interaction and hence this diagram contributes solely to $u_v$ at $x=1$.
$P_1^{p,a}=F_{1p}^a(Q^2=0)$, is the kindred axial-vector diquark contribution.  At $x=1$ this contributes twice as much to $d_v$ as it does to $u_v$.
$P_1^{p,m}=F_{1p}^m(Q^2=0)$, is the contribution to the proton's charge arising from diagrams with a different diquark component in the initial and final state.  The existence of this contribution relies on the exchange of a quark between the diquark correlations and hence it contributes twice as much to $u_v$ as it does to $d_v$.  If one uses the ``static approximation'' to the nucleon form factor, as with the treatment of the contact-interaction in Ref.\,\cite{Wilson:2011rj}, then $P_1^{p,m}\equiv 0$.  It is plain from Eq.\,\eqref{dvuvF1result} that $d_v/u_v=0$ in the absence of axial-vector diquark correlations; i.e., in scalar-diquark-only models of the nucleon, which were once common and, despite their weaknesses, still too often employed.

Using the probabilities presented in Refs.\,\cite{Wilson:2011rj,Cloet:2008re}, one obtains:
\begin{equation}
\label{compdvonuv}
\begin{array}{l|ccccc}
 & P_1^{p,s} & P_1^{p,a} & P_1^{p,m} & \frac{d_v}{u_v} & \frac{F_2^n}{F_2^p} \\\hline
M(p^2) & 0.60 & 0.25 & 0.15 & 0.28 & 0.49\\
%
%
\mbox{M=constant} &  0.78 & 0.22 & 0\rule{1.2em}{0ex} & 0.18 & 0.41\\
\end{array}\;,
\end{equation}
Both rows in Eq.\,\eqref{compdvonuv} are consistent with $d_v/u_v= 0.23\pm 0.09$ ($F_2^n/F_2^p = 0.45 \pm 0.08$) inferred recently via consideration of electron-nucleus scattering at $x>1$ \cite{Hen:2011rt}.  On the other hand, this is also true of the result obtained through a naive consideration of the isospin and helicity structure of a proton's light-front quark wave function at $x\sim 1$, which suggests that $d$-quarks are five-times less likely than $u$-quarks to possess the same helicity as the proton they comprise; viz., $d_v/u_v=0.2$ \cite{Farrar:1975yb}.  Plainly, contemporary experiment-based analyses do not provide a particularly discriminating constraint.  Future experiments with a tritium target should help \cite{Holt:2010zz}, emphasising again the critical interplay between experiment and theory in elucidating the nature of the strong interaction.

\section{Epilogue}
\label{sec:Epilogue}
QCD is the most interesting part of the Standard Model and Nature's only example of a truly nonperturbative fundamental theory.  Whilst confinement remains a puzzle, dynamical chiral symmetry breaking (DCSB) is a fact.  It is manifest in dressed-propagators and vertices, and, amongst other things, responsible for:
the transformation of the light current-quarks in QCD's Lagrangian into heavy constituent-like quarks;
the unnaturally small values of the masses of light-quark pseudoscalar mesons and the $\eta$-$\eta^\prime$ splitting;
the unnaturally strong coupling of pseudoscalar mesons to light-quarks -- $g_{\pi \bar q q} \approx 4.3$;
and the unnaturally strong coupling of pseudoscalar mesons to the lightest baryons -- $g_{\pi \bar N N} \approx 12.8 \approx 3 g_{\pi \bar q q}$.

These notes highlight the dramatic impact that DCSB has upon observables in hadron physics.  A distinctive marker for DCSB is the behaviour of the dressed-quark mass function.  The momentum dependence manifest in Fig.\,\ref{fig:Fig1} is an essentially quantum field theoretical effect.  Exposing and elucidating its consequences therefore requires a nonperturbative and symmetry-preserving approach, where the latter means preserving Poincar\'e covariance, chiral and electromagnetic current-conservation, etc.  The Dyson-Schwinger equations (DSEs) provide such a framework.

This is an exciting time in hadron physics.  These notes emphasise one of the reasons.  Namely, through the DSEs, one is unifying phenomena as diverse as: the hadron spectrum; hadron elastic and transition form factors; parton distribution functions; the physics of hadrons containing one or more heavy quarks; and properties of the quark gluon plasma.  The key is an understanding of both the basic origin of visible mass and the far-reaching consequences of the mechanism responsible; i.e., DCSB.  Through continuing feedback between experiment and theory, these studies should lead to an explanation of confinement, the phenomenon that makes nonperturbative QCD the most interesting piece of the Standard Model.  They might also provide an understanding of nonperturbative physics that enables the formulation of a realistic extension of that model.

\bibliographystyle{../z10KITPC/h-physrev4}
\bibliography{../z10KITPC/CollectiveKITPC}

\begin{thebibliography}{10}

\bibitem{Krein:1990sf}
C.~D. Roberts, A.~G. Williams and G.~Krein,
\newblock Int. J. Mod. Phys. {\bf A7}, 5607 (1992).

\bibitem{Roberts:2007ji}
C.~D. Roberts,
\newblock Prog. Part. Nucl. Phys. {\bf 61}, 50 (2008).

\bibitem{Holt:2010vj}
R.~J. Holt and C.~D. Roberts,
\newblock Rev. Mod. Phys. {\bf 82}, 2991 (2010).

\bibitem{Chang:2011vu}
L.~Chang, C.~D. Roberts and P.~C. Tandy,
\newblock Chin. J. Phys. {\bf 49}, 955 (2011).

\bibitem{Bashir:2012}
A.~Bashir {\em et~al.},
\newblock (arXiv:1201.3366 [nucl-th]).

\bibitem{Nambu:1961tp}
Y.~Nambu and G.~Jona-Lasinio,
\newblock Phys.Rev. {\bf 122}, 345 (1961).

\bibitem{Bhagwat:2003vw}
M.~Bhagwat, M.~Pichowsky, C.~Roberts and P.~Tandy,
\newblock Phys.Rev. {\bf C68}, 015203 (2003).

\bibitem{Bhagwat:2006tu}
M.~S. Bhagwat and P.~C. Tandy,
\newblock AIP Conf. Proc. {\bf 842}, 225 (2006).

\bibitem{Bowman:2005vx}
P.~O. Bowman {\em et~al.},
\newblock Phys. Rev. {\bf D71}, 054507 (2005).

\bibitem{Casher:1974xd}
A.~Casher and L.~Susskind,
\newblock Phys. Rev. {\bf D9}, 436 (1974).

\bibitem{MichelsonMorley}
A.~A. Michelson and E.~W. Morley,
\newblock American Journal of Science {\bf 34}, 333 (1887).

\bibitem{Turner:2001yu}
M.~S. Turner,
\newblock (astro-ph/0108103).

\bibitem{Brodsky:2009zd}
S.~J. Brodsky and R.~Shrock,
\newblock Proc. Nat. Acad. Sci. {\bf 108}, 45 (2011).

\bibitem{Brodsky:2010xf}
S.~J. Brodsky, C.~D. Roberts, R.~Shrock and P.~C. Tandy,
\newblock Phys. Rev. {\bf C82}, 022201(R) (2010).

\bibitem{Chang:2011mu}
L.~Chang, C.~D. Roberts and P.~C. Tandy,
\newblock Phys. Rev. {\bf C} (Rapid Comm., \emph{in press}).

\bibitem{Brodsky:2008be}
S.~J. Brodsky and R.~Shrock,
\newblock Phys. Lett. {\bf B666}, 95 (2008).

\bibitem{Glazek:2011vg}
S.~D. Glazek,
\newblock Acta Phys. Polon. {\bf B42}, 1933 (2011).

\bibitem{Weinberg:1978kz}
S.~Weinberg,
\newblock Physica {\bf A96}, 327 (1979).

\bibitem{Maris:1997hd}
P.~Maris, C.~D. Roberts and P.~C. Tandy,
\newblock Phys. Lett. {\bf B420}, 267 (1998).

\bibitem{Maris:1997tm}
P.~Maris and C.~D. Roberts,
\newblock Phys. Rev. {\bf C56}, 3369 (1997).

\bibitem{Langfeld:2003ye}
K.~Langfeld {\em et~al.},
\newblock Phys. Rev. {\bf C67}, 065206 (2003).

\bibitem{GellMann:1951rw}
M.~Gell-Mann and F.~Low,
\newblock Phys. Rev. {\bf 84}, 350 (1951).

\bibitem{Alkofer:1993gu}
R.~Alkofer, A.~Bender and C.~D. Roberts,
\newblock Int. J. Mod. Phys. {\bf A10}, 3319 (1995).

\bibitem{Flambaum:2007mj}
V.~V. Flambaum and R.~B. Wiringa,
\newblock Phys. Rev. {\bf C76}, 054002 (2007).

\bibitem{Courtland:2010zz}
R.~Courtland,
\newblock New Sci. {\bf 207N2776}, 10 (2010).

\bibitem{Andersen:2011yj}
J.~R. Andersen {\em et~al.},
\newblock Eur. Phys. J. Plus {\bf 126}, 81 (2011).

\bibitem{Davydychev:2000rt}
A.~I. Davydychev, P.~Osland and L.~Saks,
\newblock Phys. Rev. {\bf D63}, 014022 (2001).

\bibitem{Chang:2010hb}
L.~Chang, Y.-X. Liu and C.~D. Roberts,
\newblock Phys. Rev. Lett. {\bf 106}, 072001 (2011).

\bibitem{Chang:2011tx}
L.~Chang, I.~C. Clo{\"e}t, C.~D. Roberts and H.~L.~L. Roberts,
\newblock AIP Conf. Proc. {\bf 1354}, 110 (2011).

\bibitem{Wilson:2011rj}
D.~J. Wilson, I.~C. Clo{\"e}t, L.~Chang and C.~D. Roberts,
\newblock (arXiv:1112.2212 [nucl-th]).

\bibitem{Jegerlehner:2009ry}
F.~Jegerlehner and A.~Nyffeler,
\newblock Phys. Rept. {\bf 477}, 1 (2009).

\bibitem{Goecke:2011pe}
T.~Goecke, C.~S. Fischer and R.~Williams,
\newblock Phys. Lett. {\bf B704}, 211 (2011).

\bibitem{Bowman:2004jm}
P.~O. Bowman {\em et~al.},
\newblock Phys. Rev. {\bf D70}, 034509 (2004).

\bibitem{Aguilar:2009nf}
A.~Aguilar, D.~Binosi, J.~Papavassiliou and J.~Rodriguez-Quintero,
\newblock Phys.Rev. {\bf D80}, 085018 (2009).

\bibitem{Boucaud:2011ug}
P.~Boucaud {\em et~al.},
\newblock (arXiv:1109.1936 [hep-ph]).

\bibitem{Qin:2011dd}
S.-x. Qin, L.~Chang, Y.-x. Liu, C.~D. Roberts and D.~J. Wilson,
\newblock Phys. Rev. {\bf C84}, 042202(R) (2011).

\bibitem{Bloch:2002eq}
J.~C.~R. Bloch,
\newblock Phys. Rev. {\bf D66}, 034032 (2002).

\bibitem{Bashir:2011dp}
A.~Bashir, R.~Bermudez, L.~Chang and C.~D. Roberts,
\newblock (arXiv:1112.4847 [nucl-th]).

\bibitem{Cloet:2008fw}
I.~C. Clo{\"e}t and C.~D. Roberts,
\newblock PoS {\bf LC2008}, 047 (2008).

\bibitem{Bhagwat:2004hn}
M.~S. Bhagwat, A.~H{\"o}ll, A.~Krassnigg, C.~D. Roberts and P.~C. Tandy,
\newblock Phys. Rev. {\bf C70}, 035205 (2004).

\bibitem{Ezawa:1974wm}
Z.~F. Ezawa,
\newblock Nuovo Cim. {\bf A23}, 271 (1974).

\bibitem{Farrar:1975yb}
G.~R. Farrar and D.~R. Jackson,
\newblock Phys. Rev. Lett. {\bf 35}, 1416 (1975).

\bibitem{Nguyen:2011jy}
T.~Nguyen, A.~Bashir, C.~D. Roberts and P.~C. Tandy,
\newblock Phys. Rev. {\bf C83}, 062201(R) (2011).

\bibitem{Hecht:2000xa}
M.~B. Hecht, C.~D. Roberts and S.~M. Schmidt,
\newblock Phys. Rev. {\bf C63}, 025213 (2001).

\bibitem{Conway:1989fs}
J.~S. Conway {\em et~al.},
\newblock Phys. Rev. {\bf D39}, 92 (1989).

\bibitem{Wijesooriya:2005ir}
K.~Wijesooriya, P.~E. Reimer and R.~J. Holt,
\newblock Phys. Rev. {\bf C72}, 065203 (2005).

\bibitem{Aicher:2010cb}
M.~Aicher, A.~Sch{\"a}fer and W.~Vogelsang,
\newblock Phys.\ Rev.\ Lett. {\bf 105}, 252003 (2010).

\bibitem{Badier:1980jq}
J.~Badier {\em et~al.},
\newblock Phys. Lett. {\bf B93}, 354 (1980).

\bibitem{Badier:1983mj}
J.~Badier {\em et~al.},
\newblock Z. Phys. {\bf C18}, 281 (1983).

\bibitem{Bicudo:2001jq}
P.~Bicudo {\em et~al.},
\newblock Phys. Rev. {\bf D65}, 076008 (2002).

\bibitem{Maris:2000sk}
P.~Maris and P.~C. Tandy,
\newblock Phys. Rev. {\bf C62}, 055204 (2000).

\bibitem{Roberts:2010rn}
H.~L.~L. Roberts {\em et~al.},
\newblock Phys. Rev. {\bf C82}, 065202 (2010).

\bibitem{Cahill:1988dx}
R.~T. Cahill, C.~D. Roberts and J.~Praschifka,
\newblock Austral. J. Phys. {\bf 42}, 129 (1989).

\bibitem{Cahill:1987qr}
R.~T. Cahill, C.~D. Roberts and J.~Praschifka,
\newblock Phys. Rev. {\bf D36}, 2804 (1987).

\bibitem{Eichmann:2011vu}
G.~Eichmann,
\newblock Phys. Rev. {\bf D84}, 014014 (2011).

\bibitem{Roberts:2011cf}
H.~L.~L. Roberts, L.~Chang, I.~C. Clo{\"e}t and C.~D. Roberts,
\newblock Few Body Syst. {\bf 51}, 1 (2011).

\bibitem{Nakamura:2010zzi}
K.~Nakamura {\em et~al.},
\newblock J. Phys. {\bf G37}, 075021 (2010).

\bibitem{Aznauryan:2007ja}
I.~G. Aznauryan,
\newblock Phys. Rev. {\bf C76}, 025212 (2007).

\bibitem{Cardarelli:1996vn}
F.~Cardarelli, E.~Pace, G.~Salme and S.~Simula,
\newblock Phys. Lett. {\bf B397}, 13 (1997).

\bibitem{Suzuki:2010yn}
N.~Suzuki, T.~Sato and T.~S.~H. Lee,
\newblock Phys. Rev. {\bf C82}, 045206 (2010).

\bibitem{JuliaDiaz:2009ww}
B.~Julia-Diaz {\em et~al.},
\newblock Phys. Rev. {\bf C80}, 025207 (2009).

\bibitem{Dugger:2009pn}
M.~Dugger {\em et~al.},
\newblock Phys. Rev. {\bf C79}, 065206 (2009).

\bibitem{Aznauryan:2009mx}
I.~Aznauryan {\em et~al.},
\newblock Phys. Rev. {\bf C80}, 055203 (2009).

\bibitem{Aznauryan:2011td}
I.~G. Aznauryan, V.~D. Burkert and V.~I. Mokeev,
\newblock (arXiv:1108.1125 [nucl-ex]).

\bibitem{Suzuki:2009nj}
N.~Suzuki {\em et~al.},
\newblock Phys. Rev. Lett. {\bf 104}, 042302 (2010).

\bibitem{Gasparyan:2003fp}
A.~M. Gasparyan, J.~Haidenbauer, C.~Hanhart and J.~Speth,
\newblock Phys. Rev. {\bf C68}, 045207 (2003).

\bibitem{Gothe:2011up}
R.~W. Gothe,
\newblock (arXiv:1108.4703 [nucl-ex]).

\bibitem{Cloet:2008re}
I.~C. Clo{\"e}t, G.~Eichmann, B.~El-Bennich, T.~Kl{\"a}hn and C.~D. Roberts,
\newblock Few Body Syst. {\bf 46}, 1 (2009).

\bibitem{Hen:2011rt}
O.~Hen, A.~Accardi, W.~Melnitchouk and E.~Piasetzky,
\newblock (arXiv:1110.2419 [hep-ph]).

\bibitem{Holt:2010zz}
R.~J. Holt and J.~R. Arrington,
\newblock AIP Conf. Proc. {\bf 1261}, 79 (2010).

\end{thebibliography}

\end{document}